# A Qualitative Comparative Study of Communication in Higher Distance Education

VASILEIOS MELLOS[1*], PETROS TRANTAS[2], KLEANTHI SANTAMOURI[3], CHRISTOS DOUVLOS[4], ALEXANDROS GAZIS[5], THEODOROS VAVOURAS[6,7]

[1] Pedagogical Training Program, ASPETE School of Pedagogical and Technological Education, 38446 Volos, GREECE

[2] Department of Primary Education, Dimokratias 1, Rhodes 85132, GREECE

[3] English Language Teacher, Private Primary Education, Athens, GREECE

[4] Information Technology Educator and Researcher in Pedagogical Psychology, Athens, GREECE

[5] Heriot-Watt University, Edinburgh EH14 4AS, UNITED KINGDOM

[6] Department of Humanities, School of Humanities, Hellenic Open University, 26335 Patras, GREECE

[7] Department of Philosophy, School of Italian Language and Literature, Aristotle University of Thessaloniki, 54124 Thessaloniki, GREECE

***Abstract*—** The rise of open and distance education has made it more important than ever to have communication tools that are simple, flexible, and good for helping students work together, talk to each other, and feel connected. Researchers have already looked at how instant messaging apps like WhatsApp and Telegram can be used for learning. Viber, on the other hand, has not been studied as much, especially when it comes to its use in higher education at a distance. This article builds on a previously published conference case study conducted at the Hellenic Open University (HOU), which examined the use of Viber in distance collaborative projects. The present study extends that work by offering a comparative discussion of communication ecosystems in higher distance education. Using ideas from connectivism learning theory, along with the concepts of social presence and community-based learning, this article looks at how chatting on Viber can add to and improve formal online learning. The findings show that Viber is not just a simple messaging app. It also works as a casual space where students pass along what they know, give each other a hand, and slowly build a sense of being part of a group. This lines up with other research showing that social media can help students in distance learning feel less isolated. Overall, the study shows why it makes sense to include informal chat platforms when planning courses for higher education at a distance.

## I. INTRODUCTION

The rapid growth of open and distance education (ODE) has changed the way higher education is delivered, [1]. It has increased access and flexibility for many learners, but it has also made issues such as interaction, collaboration, and social connectedness more important. One of the most common challenges in ODE is the feeling of isolation experienced by students, [2]. This problem has often been linked to lower engagement and reduced persistence in learning, [3], [4]. For this reason, researchers have paid increasing attention to digital communication tools that operate outside formal learning management systems (LMS) and allow students to communicate in more direct and socially meaningful ways, [5], [6].

In this setting, students often turn to social media and instant messaging apps as casual ways to stay in touch. Earlier research has pointed out the learning benefits of platforms like Facebook, WhatsApp, Telegram, and Skype, especially when it comes to helping students interact with one another, swap information, and learn together, [7], [8], [9], [10]. Students tend to like these tools because they are already used to them, they are easy to open up, and they slot right into the way students chat with each other day to day.. That said, research also points out some downsides, like getting distracted, having a harder time focusing, and the fact that these apps are not always built into lessons in a clear way, [11], [12].

More recent research has moved beyond the study of individual tools and has focused instead on the broader idea of communication ecosystems, [13]. This approach examines how formal and informal platforms work together and shape the learning experience in distance education, [14], [15], [16]. From this perspective, learning is not limited to the official digital classroom or the institutional LMS. Rather, it is formed through social and technological connections, where students exchange information, ask for support, and build meaning through participation in wider digital networks. This view is closely related to connectivist learning theory, which understands knowledge as something that emerges through connections among people, digital tools, and information sources, [17], [18].

Although there is now a growing body of research on the use of IM applications in education, Viber remains relatively underexamined in the international literature, especially in higher distance education. Existing studies have mainly approached Viber as a tool for communication efficiency or social interaction, rather than as part of collaborative learning practices, [19], [20]. In the Greek context, and particularly at the Hellenic Open University (HOU), students appear to use

Viber extensively as an informal tool for coordination, peer support, and everyday communication. However, systematic research on its pedagogical role remains, [21].

Building on these observations, the present article examines Viber within the wider communication ecosystem of higher distance education. It places Viber alongside other communication practices reported in previous research and compares its role with that of other social media and IM platform, [22], [23]. Despite the increasing use of such platforms, there is still limited comparative research on the specific affordances of Viber within the Greek distance education context. To help fill this gap, the study is built around the following research question: How does using Viber as an informal communication space add to formal learning and help support group learning and social connection in higher distance education?

The goal of this study is to look at how chatting on Viber helps students learn together, feel socially present, and develop a sense of belonging in distance learning. It also examines how these functions compare with those identified in studies of other social media and IM platforms. By adopting a a clearer picture of how everyday communication tools play a key role in shaping learning in higher distance education.

## II. Theoretical Framework

This study is built on a framework that pulls together three ideas: connectivist learning theory, social presence theory, and the concept of communication ecosystems. Looking at these together gives us a helpful way to understand how everyday communication tools, like instant messaging apps, can support learning in open and distance education.While each of these three ideas focuses on something different about online learning, they all point to the same problem in higher distance education: the gap between learning that is run through an LMS and learning that students drive themselves through their own networks. Learning management systems are usually built around set content, control from the institution, and planned interaction. Because of this, they tend to follow more old-school ideas about how teaching and learning should work.

### *A. Connectivist Learning Theory*

Connectivism views learning as a networked process in which knowledge is distributed across connections among people, digital tools, and information source, [17], [18], [24], [25]. From this point of view, learning is not just about taking in content. It also depends on whether students can build, keep up, and move around within their own networks. Digital tools are not just ways to pass along information. They are an active part of the learning setting, since they shape how people find, share, and build knowledge.

This idea is especially important in open and distance education, where students rely a lot on digital communication because they are not in the same room together. Instant messaging apps can help students swap information quickly, give each other feedback, and stay in touch all the time. These features line up well with the main ideas behind connectivism, like independence, openness, having a mix of viewpoints, and being able to interact with others, [17], [18]. Previous research has shown that networked communication tools can support informal learning and peer-based knowledge construction, particularly when learners organize themselves around common academic goals, [16], [18].

From a connectivist perspective, the educational value of platforms such as Viber does not lie in predefined instructional functions. Rather, it lies in their ability to sustain learning networks that extend beyond the formal boundaries of the institution. Through these networks, learners can exchange knowledge, draw on collective experience, and access support from their peers. In this way, learning becomes a continuous and socially embedded process.

### *B. Social Presence and Community in Distance Education*

Social presence theory, which was expanded through the Community of Inquiry (CoI) framework, [3], focuses on the extent to which learners perceive others as “real” in mediated communication environments. In distance education, social presence is seen as a big part of what keeps students engaged, happy with their studies, and doing well academically, [14], [15].

Research has shown that signs of social presence, like making messages feel personal, showing emotion, going back and forth in conversation, and noticing your classmates, are linked to better results in online learning, [14], [15], [16]. Instant messaging apps can boost social presence because they let people talk in real time or close to it. They also make room for casual language, quick replies, and steady back-and-forth, all of which can make conversations feel warmer and more direct.

In higher distance education, having a stronger sense of social presence can help students feel less alone and make it easier for learning communities to grow, [21]. Casual chat spaces often become places where students ask questions, double-check assignments, share study tips, and cheer each other on. These things matter even more in places like the Hellenic Open University, where many students are juggling their studies with work and family life.

### *C. Communication Ecosystems in Higher Distance Education*

The concept of communication ecosystems moves the analysis beyond individual platforms and focuses on how different formal and informal communication tools work together within a learning environment, [14], [15], [16]. Rather than treating learning management systems, social media, and instant messaging applications as separate tools, this approach views them as interconnected parts of a socio-technical system.

Research on communication ecosystems shows that students move between platforms depending on their purpose, the urgency of communication, and the social norms of each space, [10], [11]. Formal LMS environments are usually used for content delivery, official announcements, assessment, and course administration. Informal tools, on the other hand, are often used for coordination, peer collaboration, quick questions, and emotional support. This division of communication functions shows that informal platforms do not necessarily replace institutional systems. Instead, they often complement them.

In this study, Viber is examined as part of the communication ecosystem of higher distance education. Its wide use in the Greek context, its ease of access, and its connection with students' everyday communication practices make it an important space for studying collaborative learning. By placing Viber within a broader ecosystemic perspective, the study aims to show how informal communication infrastructures can support learning, social presence, and community formation in ODE settings.

Earlier research has found that social media and messaging apps usually run side by side with the official tools used by schools, creating a mixed setup rather than fully separate or competing systems, [10], [11]. Students pick different apps for different reasons, like needing a quick reply, looking for emotional support, sorting out coursework, or just feeling that one app fits better socially. This backs up the idea that digital learning environments have different tools doing different jobs.

### *D. From LMS-Centered Instruction to Networked Learning Logics*

Connectivist learning offers a different view from LMS-centered instruction. It emphasizes distributed knowledge, horizontal interaction, and the role of digital technologies as dynamic nodes within learning networks, [17], [18], [24], [25]. From this perspective, learning extends beyond formal platforms and develops through ongoing connections among learners, tools, and information sources.

Communication ecosystem research supports this distinction by showing that students often assign different communication functions to different platforms. Formal systems are mainly used for administrative and academic purposes, while informal tools are used for collaboration, coordination, and social presence, [10], [14], [15].

In the present study, this theoretical distinction provides the basis for interpreting the pedagogical role of Viber. Viber is understood as a networked communication space that complements, rather than replaces, the institutional LMS. To clarify the operational and pedagogical differences between these two approaches, Table I compares LMS-centered logic with connectivist learning logic. The comparison shows how the shift from a centralized and hierarchical system to a distributed and networked environment can change the dynamics of interaction, knowledge sharing, and student engagement.

TABLE I. LMS-CENTERED LOGIC VS. CONNECTIVIST LEARNING LOGIC

| Dimension | LMS-centered logic | Connectivist logic |
|---|---|---|
| Knowledge flow | Structured and mainly linear | Non-linear and emergent |
| Learning space | Institutional LMS | Networked digital environments |
| Learner role | Content consumer | Active network participant |
| Technology role | Delivery and management | Connectivity and interaction |
| Communication authority | Hierarchical | Peer-driven |
| Interaction | Scheduled and course-bound | Continuous and emergent |
| Social presence | Peripheral | Central |
| Adaptability | Limited | High |
| Study example | LMS forums | Viber groups |

## III. METHODOLOGY

This study adopts a qualitative comparative case study design to examine the role of instant messaging applications as part of communication ecosystems in higher distance education. The methodological approach builds on and extends the empirical findings of a previously published case study conducted at the Hellenic Open University, [26].

Rather than treating the HOU case as an isolated example, the study uses a longitudinal and comparative analytical perspectiveIt pulls together first-hand data from the Hellenic Open University setting with a careful review of international research published between 2015 and 2025. By drawing on a mix of sources, including structured observation, content analysis, and reflections from the participants themselves, the study is able to cross-check its findings from several angles.

This way of working makes it possible to get a deeper look at how chatting on Viber changes and grows across the different stages of group work.. It also makes it possible to move beyond a narrow view of Viber as a tool for simple information exchange and to examine its role in the formation of a socially driven, connectivist learning environment.

### *A. Research Design*

A case study methodology was selected because it allows for an in-depth examination of complex social and technological phenomena within their real-life context, [24], [25], [26]. This design is suitable for studying informal learning practices that emerge outside formal institutional structures, such as the use of Viber by distance learners.

The study focuses on how communication practices develop within a specific educational setting. It does not aim to produce statistical generalizations. Instead, it seeks to provide a detailed understanding of how informal digital communication supports collaboration, peer interaction, and social presence in higher distance education.

The comparative dimension of the study is analytical rather than experimental. The empirical findings from the HOU case are compared with findings reported in previous studies on instant messaging and social media use in higher and distance education (e.g., [10], [19], [20], [26]). This comparison helps identify similarities and differences across educational contexts and digital platforms.

### *B. Research Context and Case Selection*

The main setting for this research is the Hellenic Open University, the only public university in Greece that focuses entirely on open and distance learning. The teaching model at HOU is built mostly around platforms students can use on their own time, like Moodle, along with the occasional live video session. While this setup gives students flexibility, it can also leave them without enough chances to talk in real time. Because of that, students often look for other easy-to-use ways to chat that help them keep up with their studies.

This study looks at one specific group of undergraduate students who took part in distance group projects. These students chose on their own to set up Viber groups so they could coordinate in real time, talk to each other, and work together on assignments.

These digital spaces were not required or formally organized by the institution. Instead, they emerged as authentic student-driven practices and operated independently of the official LMS.

This autonomy makes the HOU context a suitable case for examining the "Third Space", [27], [28], where informal social interaction meets formal academic requirements. In this space, students can mix working on schoolwork together with giving each other support, sorting out the small day-to-day stuff, and just chatting like normal.

The group taking part was made up of adult learners from a wide range of academic and work backgrounds, which is something you often see in the HOU student body. This mix of people gives a great setting to look at how casual chat spaces can help keep a group together, support teamwork, and help students cope in the demanding world of distance learning.

### *C. Data Collection*

To ensure analytical depth and credibility, data collection followed a qualitative multi-source approach. This made it possible to triangulate findings across different types of communication and participant experience. The primary data sources included structured observation, content analysis, and participant reflective inputs.

Structured observation involved the systematic monitoring of interactions within Viber groups created for academic collaboration. This process focused on the frequency of participation and on the main functions of communication, such as task-oriented interaction, social support, and coordination.

Content analysis involved a detailed examination of message exchanges. Messages were sorted by what they were trying to do, such as asking academic questions, giving each other feedback, sharing study materials, working out logistics, and offering support.Reflections shared by the participants added more depth to the picture, giving extra insight into what students went through during and after the group projects.

Similarly, participant reflective inputs provided additional qualitative insight into students' experiences during and after the collaborative projects. These reflections helped capture students' perceptions of Viber's usefulness and its role in their broader learning experience.

Data was collected throughout the whole time the group projects were running. Looking at it over this longer stretch made it possible to see how the way students interacted shifted along the way, from when the groups first came together to when the projects wrapped up.

### *D. Data Analysis*

A mixed approach was used, blending codes that came up from the data itself with categories that were already laid out in the theory behind the study. The analysis was shaped by the main ideas of connectivism, signs of social presence, and the different roles communication ecosystems play, [3], [5], [14], [15], [17], [18].

The first round of coding looked for patterns that kept showing up, like sharing knowledge, helping each other out, sorting out logistics, and showing emotion in messages.. These codes were then organized into broader themes that reflected the pedagogical and social affordances of Viber. Particular attention was given to the role of Viber as a "Third Space" for academic engagement, [28].

To strengthen the comparative dimension of the study, a cross-case thematic mapping was conducted. The themes that came out of the HOU case were placed side by side with findings from earlier studies published between 2015 and 2025 on how messaging apps and social media are used in higher education. Putting these together made it easier to spot patterns that show up across the board in digital communication, as well as the things that stand out as specific to the Greek distance education setting.

### *E. Ethical Considerations*

Ethical principles were followed all the way through the study. Taking part was up to each student, and everyone was told upfront what the research was about. All the data was kept anonymous to protect privacy, and nothing that could identify a specific person was used in the analysis.

The study also made sure to respect the casual, student-run nature of the Viber groups. The analysis stuck to looking at how people communicated and how that tied into learning, without putting individual students or their private business on display. Overall, the study stuck to the usual ethical rules for education research and made sure students' privacy and freedom to choose were looked after.

## IV. Findings

The findings of this study are built around the first line of analysis, which looks at how different communication ecosystems in higher distance education compare with each other. This axis brings together empirical observations from the HOU case study and findings from previous research. The aim is to identify both common patterns and distinctive features of communication mediated by instant messaging applications.

By examining key dimensions such as authority, connectivity, social presence, and the flow of information, the analysis provides a clearer understanding of how Viber functions as a "third space" in distance education. In this role, Viber bridges formal institutional requirements with the more fluid, immediate, and socially embedded needs of a learning community.

### *A. Viber as a Social and Collaborative Space*

Consistent with earlier studies on the communicative affordances of instant messaging application, [19], [20], the findings indicate that Viber plays an important role in supporting rapid information exchange among distance learners. However, the analysis of the HOU case shows that its use goes beyond functional communication.

Viber works as a shared space for both socializing and working together, where students sort out what their assignments mean, plan out group work, pass along useful resources, and stay connected with one another throughout the projects. In other words, the app supports both the practical side of learning and the social side of it.

This finding is consistent with the view that communication technologies in distance education should be examined as socio-technical environments rather than as neutral tool, [8]. In the HOU context, Viber-supported interaction helped students develop a stronger sense of belonging and collaboration. This matters a lot in learning settings where people are not online at the same time, since the social side of things can be weaker or harder to keep going. Unlike formal LMS platforms, where messages tend to be focused on tasks, take a while to come through, and are usually started by the instructor, Viber let students kick off their own conversations and keep them going all the time. Because of this, it helped make the learning experience feel more connected and tight-knit.

### *B. Communication Ecosystems and Social Presence*

From a communication ecosystem perspective, [14], [15], Viber can be understood as one part of a broader network of platforms that support learning in different ways. The findings indicate a clear division of communicative functions. Formal platforms were mainly used for content access, official announcements, and assessment. Viber, by contrast, was used for coordination, clarification of academic tasks, peer support, and the expression of social presence.

Several indicators of social presence were evident in Viber interactions. These included informal language, acknowledgment of peers, quick responses, emotional support, and continuous participation. Such indicators are similar to those identified by [14], [15], [16], as factors associated with successful online learning interactions.

Although the present study did not directly measure academic performance, the sustained engagement observed in the Viber groups suggests that the platform may support learner persistence and collaborative effectiveness. In particular, Viber appeared to help students remain connected to both the academic task and the learning community.

### *C. Towards a Comparative Understanding of Informal Communication Tools*

Looking at things side by side shows both the things Viber has in common with other messaging or social media apps talked about in recent studies, like WhatsApp, Facebook, and Telegram, and the things that set them apart. Earlier research has shown again and again that casual chat tools can help students work together, make distance learners feel less alone, and keep students more involved in their courses, [16], [26].

The HOU case adds something specific about its own setting to the conversation. In Greek distance education, Viber seems to play an especially big role because it is already woven into the way students talk to each other every day. Because it is so familiar, students find it easy to jump in and stay active in the group over time.

All in all, the findings show that Viber sits as a key piece within the communication setup of higher distance education. Its job is not just to pass information back and forth. It also helps students build learning networks together, strengthens friendships between classmates, and builds up a sense of community among distance learners. For this reason, casual apps like Viber should be seen as meaningful "third spaces" for learning, not just treated as outside extras or backup communication tools.

To further illustrate these distinctions, Table II provides a comparative overview of the functional and pedagogical affordances of Viber in relation to formal LMS platforms and other instant messaging applications.

TABLE II. COMPARATIVE OVERVIEW OF COMMUNICATION PLATFORMS IN HIGHER DISTANCE EDUCATION

| **Dimension** | **Formal LMS, e.g., Moodle** | **Instant messaging apps, general** | **Viber, HOU case study** |
|---|---|---|---|
| Primary function | Content delivery, assessment, and official announcements | Informal communication and coordination | Informal academic collaboration and social interaction |
| Communication structure | Instructor-centered and hierarchical | Peer-centered and networked | Peer-centered and highly networked |
| Interaction rhythm | Mostly asynchronous and often delayed | Synchronous or near-synchronous | Continuous and near-synchronous |
| Social presence | Limited and mainly task-oriented | Moderate | High, through emotional support, immediacy, and belonging |
| Student initiative | Low to moderate | High | Very high |
| Integration into daily practices | Low | Medium to high | Very high, as it is embedded in everyday communication |
| Role in the learning ecosystem | Core but relatively rigid | Supplementary | Complementary and socially important |

## V. DISCUSSION AND COMPARISON WITH PRIOR EMPIRICAL STUDIES

The purpose of this study was to contribute to a comparative understanding of communication ecosystems in higher distance education by examining the role of Viber as an informal, but pedagogically meaningful, communication environment. The discussion interprets the findings in relation to previous empirical studies, connectivist learning theory, and the specific conditions of ODE.

The educational value of Viber identified in this study is consistent with the findings of [19], [20], who reported that Viber supports efficient information sharing and social interaction among university students. Their study mainly focused on the social dimension of Viber use. The present study extends this view by showing how social interaction can also support collaborative learning and networked knowledge construction.

The findings are also in line with the study by [10], which showed that Romanian university students used Viber mainly for personal communication and informal coordination. The HOU case backs this pattern up, but it also shows that casual use can become a real part of teaching and learning in distance education. In this setting, chatting on Viber acted as a kind of support system for working together on schoolwork and helped fill in the gaps in communication that the school's own platforms did not fully cover.

This lines up with research on casual learning spaces at the Hellenic Open University, [21], showing that students go looking for other ways to communicate when the official tools do not give them enough room to interact. Viber groups popped up naturally as something students started on their own to deal with the challenges of distance learning, especially the need for quick replies, emotional support, and help from classmates.

### A. Viber within a Connectivist Learning Framework

When you look at the findings through the lens of connectivism, [17], [18], [24], [25], they suggest that Viber works as a point in a bigger learning network, rather than being just a standalone chat app on its own. Building knowledge did not only happen through course content or activities led by the instructor. It also grew out of students talking to each other, getting quick feedback, working through problems together, and being involved all the time. This fits with the connectivist idea that learning is spread out across both social and tech networks.

Unlike formal learning management systems, which often organize interaction in a more hierarchical way, Viber supported horizontal communication among learners. Participants were not only receivers of information. They also acted as information seekers, content contributors, peer supporters, and coordinators of group activity.

Similar patterns have been observed in studies of WhatsApp- and Telegram-supported learning, [16], [26]. However, the present study adds to this literature by showing how Viber-supported networks can sustain collaboration over time in the specific context of the HOU.

### B. Informal Communication and Social Presence in ODE

A key contribution of this study is that it shows how informal communication tools can strengthen social presence in distance education. The indicators identified in Viber interactions included immediacy, informal language, emotional support, frequent acknowledgment of peers, and continuous responsiveness. These indicators correspond with established dimensions of social presence in online learning environments, [3], [14].

Consistent with previous research at the HOU, [21], the findings confirm that students use external tools to compensate for the structural limitations of formal platforms. However, this study moves beyond simply documenting this practice. It shows that Viber-mediated interaction can contribute to a sense of belonging and shared action, helping to reduce the isolation that is often reported in distance education literature, [27], [28], [29].

In this sense, Viber operates as a form of social support within the learning process. It helps connect students who might otherwise experience their studies as fragmented or individualised.

### C. Comparative Positioning within the Literature

From a comparative perspective, the findings confirm and extend existing research on the role of instant messaging applications (IMAs) in higher education. The authors in [19], [20], presented Viber mainly as a medium for rapid information exchange, while [10], described its use as mostly personal and informal. The present study brings these two perspectives together. It shows that informality does not necessarily weaken academic learning. In distance education, it may actually support learning by creating conditions for peer help, emotional reassurance, and practical coordination.

In comparison with international literature that focuses mainly on WhatsApp or Facebook, [16], [26], Viber appears to have a strong socio-technical fit within the Greek educational context. Its widespread use in everyday communication creates a very low technological barrier for students. This allows communication to move more easily between personal, professional, and academic contexts.

This contextual embeddedness supports sustained engagement and also shows why communication platforms should not be evaluated only in technical terms. Their educational role depends on the cultural, institutional, and social context in which they are used.

Earlier studies, such as [8], [9], [10], highlighted the influence of WhatsApp on student performance in tertiary education. The present study focuses instead on how Viber functions as a pedagogical bridge within the Greek distance learning context. To synthesize these distinctions, Table III presents a comparative analysis of findings from previous studies and the present research

TABLE III. COMPARISON OF FINDINGS FROM PREVIOUS STUDIES AND THE PRESENT RESEARCH.

| Study | Context | Main focus | Key findings | Significance for the present study |
|---|---|---|---|---|
| [10] | University students, Romania | Personal and informal use of Viber | Viber was used mainly for social purposes, with limited academic intent. | Shows how informal communication can gradually acquire pedagogical value in distance education. |

| Study | Context | Main focus | Key findings | Significance for the present study |
|---|---|---|---|---|
| [21] | HOU students, Greece | Use of external digital tools | Students actively used tools beyond official institutional platforms. | Supports the view that Viber can operate as a concrete informal tool within the Greek ODE context. |
| [19] | Higher education | Information exchange through Viber | Viber improved communication efficiency and speed. | Extends the focus from information exchange to collaboration, peer support, and belonging. |
| [30] | Tertiary education | WhatsApp and learning opportunities | Messaging apps can support motivation by meeting students' psychological and social needs. | Links close, informal messaging practices with student persistence in demanding distance learning programmes. |
| [21] | HOU postgraduate students | Social media and isolation | Social tools helped reduce feelings of isolation in distance learning. | Connects reduced isolation with sustained collaborative practices. |
| [31] | Online postgraduate courses | Messaging apps and LMS use | Mobile messaging reduced anxiety and helped students ask questions more confidently. | Positions Viber as a safer space for negotiating academic meaning and uncertainty. |
| [32] | Higher education, meta-analysis | Affordances of mobile instant messaging apps | Messaging apps showed strong potential for engagement and interaction compared with LMS forums. | Supports the role of Viber in real-time interaction and peer-based knowledge construction. |
| [33] | University students | Viber as a support tool | Viber improved students' attitudes toward complex academic tasks, such as projects. | Shows how Viber can act as a scaffold for the completion of collaborative academic projects. |
| [26] | HOU undergraduate students | Viber in collaborative projects | Viber supported coordination and peer-to-peer assistance in group work. | Repositions the local HOU findings within a broader comparative and theoretical framework. |
| [34] | Online education | Peer socialisation through external apps | Tools outside the LMS provided important "just-in-time" support that email could not always offer. | Positions Viber as a socially supportive platform for timely peer interaction. |

## VI. Conclusions

This article looked at communication ecosystems in higher distance education, with a particular focus on the role Viber plays in teaching and learning at the Hellenic Open University. By bringing together evidence from a case study with findings from earlier research, the study shows that casual communication apps are not just side tools or extras. Instead, they form part of the everyday learning infrastructure of open and distance education.

Instant messaging applications are now deeply connected with students' daily communication practices. In higher distance education, these tools are often used not only for personal communication but also for academic purposes. The findings of this study indicate that Viber functions as a low-threshold digital space that supports quick information exchange, peer coordination, and social presence beyond the formal learning management system.

The analysis suggests that Viber is more than a simple communication tool. It works as a socially embedded learning space where students collaborate, support one another, and develop a sense of belonging. When compared with previous studies on instant messaging applications in higher education, the findings offer a clearer understanding of how informal communication platforms can enrich distance learning environments. For this reason, such platforms should be seen as meaningful parts of contemporary communication ecosystems in higher distance education.

The study also confirms findings from earlier research showing that instant messaging applications support rapid information exchange, peer coordination, and continuous connectivity in higher education, [10], [19], [20]. However, the present article extends this discussion by showing how Viber also supports collaboration, social presence, and community formation among distance learners. In this respect, the study moves beyond a narrow tool-based evaluation and focuses on the wider socio-pedagogical role of communication technologies in ODE.

From a theoretical perspective, the findings support the connectivist view that learning develops through networks of people, technologies, and information sources, [17], [18], [24], [25]. In the HOU case, Viber acted as an accessible node within this wider learning network. It enabled horizontal knowledge exchange, learner-driven interaction, and peer support, while also complementing the formal LMS. The study also contributes to communication ecosystem research by showing how students use different platforms for different communicative purposes, depending on their affordances, social meaning, and practical usefulness, [14], [15].

The practical takeaways matter for any school that offers distance education. Universities should keep in mind that casual chat spaces can help learning, especially in areas where the official platforms tend to fall short, like fast responses, emotional support, and feeling close to classmates. That does not mean schools should take these tools over or fold them into their official systems. Instead, teachers might just need to notice the role these apps play and think about how they can back them up in ways that are fair and thoughtful, while still letting students do their own thing. Giving these tools that kind of recognition could help students work together more, stick with their studies, and feel more connected as a group, [27].

That said, the study does have its limits. Since it is a qualitative case study set in one specific school and cultural setting, you cannot turn its findings into broad statistical claims. Future research could look at similar practices at other universities, bring in numbers about learning outcomes, or line up several messaging apps like Viber, WhatsApp, and Telegram to see how they compare in distance education. More research is also needed on how casual communication tools tie into student grades and whether students keep going with their studies over the long run.

To wrap up, this study offers both real-world evidence and a theory-based way of seeing messaging apps as meaningful parts of communication ecosystems in higher distance education. By focusing on Viber, the article shows how informal digital practices can support collaborative learning, social connectedness, and student persistence. More broadly, students' preference for Viber points to an important pedagogical shift: learning value is increasingly created in spaces where social presence, immediacy, and peer interaction come together, and not only through formal instruction, [35].

## References

[1] Qayyum, A., & Zawacki-Richter, O. (2019). The state of open and distance education. In *Open and distance education in Asia, Africa and the Middle East: National perspectives in a digital age* (pp. 125-140). Singapore: Springer Singapore. https://doi.org/10.1007/978-981-13-5787-9_14

[2] Thacker, I., Seyranian, V., Madva, A., Duong, N. T., & Beardsley, P. (2022). Social connectedness in physical isolation: Online teaching practices that support under-represented undergraduate students' feelings of belonging and engagement in STEM. *Education Sciences*, *12*(2), 61. https://doi.org/10.3390/educsci12020061

[3] Garrison, D. R., Anderson, T., & Archer, W. (1999). Critical inquiry in a text-based environment: Computer conferencing in higher education. *The internet and higher education*, *2*(2-3), 87-105. https://doi.org/10.1016/S1096-7516(00)00016-6

[4] Kostopoulou, E., & Kostopoulos, K. P. (2023). Distance education and digital technologies in adult education during the COVID-19 pandemic: Views of postgraduate students in Hellenic Open University, Greece. *European Journal of Alternative Education Studies*, *8*(3). https://oapub.org/edu/index.php/ejae/article/view/5042 [Access Date: 01/01/2026]

[5] Simelane-Mnisi, S. (2023). Effectiveness of LMS digital tools used by the academics to foster students' engagement. *Education Sciences*, *13*(10), 980. https://doi.org/10.3390/educsci13100980

[6] Zahra, O. F., Amel, N., Soufiane, O., & Mohamed, K. (2024). From platforms to online communication tools. *DIROSAT: Journal of Education, Social Sciences & Humanities*, *2*(3), 130-147. https://doi.org/10.58355/dirosat.v2i3.68

[7] Sheridan, B. J., Patel, D., Al-Bahrani, A., & Ducking, J. (2026). Students' perspectives of social media use in economics courses. *The Journal of Economic Education*, *57*(1), 39-52. https://doi.org/10.1080/00220485.2025.2554640

[8] Kent, M., & Leaver, T. (2014). *An education in Facebook?*. Croydon: Routledge. https://dl.acm.org/doi/abs/10.5555/2675380 [Access Date: 01/01/2026]

[9] Yeboah, J., & Ewur, G. D. (2014). The impact of WhatsApp messenger usage on students performance in Tertiary Institutions in Ghana. *Journal of Education and practice*, *5*(6), 157-164. https://www.iiste.org/Journals/index.php/JEP/article/view/11241 [Access Date: 01/01/2026]

[10] Voicu, M. C., & Muntean, M. (2023). Factors that influence mobile learning among university students in Romania. *Electronics*, *12*(4), 938. https://doi.org/10.3390/electronics12040938

[11] Manca, S., & Ranieri, M. (2016). Facebook and the others. Potentials and obstacles of social media for teaching in higher education. *Computers & education*, *95*, 216-230. https://doi.org/10.1016/j.compedu.2016.01.012

[12] Alamuri, S., & Miryala, R. K. (2025). Social Learning 2.0: Harnessing Digital Networks as Learning Tools in Education. *Available at SSRN 5659550*. https://papers.ssrn.com/sol3/papers.cfm?abstract_id=5659550 [Access Date: 01/01/2026]

[13] McDougall, J. (2025). *Media Literacy for the Communication Ecosystem: A Theory of Change for a Healthier Future*. Springer. https://doi.org/10.1007/978-3-032-04024-4

[14] Joksimović, S., Gašević, D., Kovanović, V., Riecke, B. E., & Hatala, M. (2015). Social presence in online discussions as a process predictor of academic performance. *Journal of Computer Assisted Learning*, *31*(6), 638-654. https://doi.org/10.1111/jcal.12107

[15] Alsayer, A. A., & Lowenthal, P. R. (2025). Measuring social presence in online learning: A validation study. *Education and Information Technologies*, *30*(5), 5655-5676. https://doi.org/10.1007/s10639-024-12972-w

[16] Kukulska-Hulme, A., & Viberg, O. (2018). Mobile collaborative language learning: State of the art. *British Journal of Educational Technology*, *49*(2), 207-218. https://doi.org/10.1111/bjet.12580

[17] Dunaway, M.K. (2011). Connectivism: Learning theory and pedagogical practice for networked information landscapes. *Reference services review*, *39*(4), 675-685. https://doi.org/10.1108/00907321111186686

[18] Duke, B., Harper, G., & Johnston, M. (2013). Connectivism as a digital age learning theory. *The International HETL Review*, *2013*(Special Issue), 4-13. https://www.semanticscholar.org/paper/Connectivism-as-a-Digital-Age-Learning-Theory-Duke-

Harper/9d499406ce42d07fc501c534eca528361ffe460f [Access Date: 01/01/2026]

[19] Shabiralyani, G., Hasan, K. S., Hamad, N., & Iqbal, N. (2015). Impact of visual aids in enhancing the learning process case research: District Dera Ghazi Khan. *Journal of education and practice*, *6*(19), 226-233. https://eric.ed.gov/?id=EJ1079541 [Access Date: 01/01/2026]

[20] Galimova, E. G., Sergeeva, O. V., Zheltukhina, M. R., Sokolova, N. L., Zakharova, V. L., & Drobysheva, N. N. (2025). Mobile learning in science education to improve higher-order thinking skills and communication skills: scoping review. *Frontiers in Communication*, *10*, 1624012. https://doi.org/10.3389/fcomm.2025.1624012

[21] Koutsogiannopoulou, N., & Manousou, E. (2024). The academic use of social media in distance education and its relation to the reduction of the feeling of isolation. *Sociol Int J*, *8*(2), 53-59. https://www.ceeol.com/search/article-detail?id=1361370 [Access Date: 01/01/2026]

[22] Mohsenian-Rad, M. (2015). Communication Ecosystem Contexts: From Mass Audiences to Mass Messages. *GSTF Journal on Media & Communications (JMC)*, *2*(2), 13. https://doi.org/10.7603/s40874-014-0013-6

[23] Palau-Sampio, D., & López-García, G. (2025). The New Communication Ecosystem. In *News, Media, and Communication in a Polarized World: A Spanish perspective* (pp. 7-14). Cham: Springer Nature Switzerland. https://doi.org/10.1007/978-3-031-86620-3_2

[24] Downes, S. (2012). Connectivism and connective knowledge: Essays on meaning and learning networks. https://uark.pressbooks.pub/edtech/chapter/connectivism-and-connective-knowledge-2/ [Access Date: 01/01/2026]

[25] Downes, S. (2023). Newer theories for digital learning spaces. In Handbook of open, distance and digital education (pp. 129-146). Singapore: Springer Nature Singapore. https://doi.org/10.1007/978-981-19-2080-6_8

[26] Ansari, J. A. N., & Khan, N. A. (2020). Exploring the role of social media in collaborative learning the new domain of learning. *Smart Learning Environments*, *7*(1), 9. https://doi.org/10.1186/s40561-020-00118-7

[27] Tait, A. W. (2018). Education for development: From distance to open education. *Journal of Learning for Development*, *5*(2). https://doi.org/10.56059/jl4d.v5i2.294

[28] Gutiérrez, K. D. (2008). Developing a sociocritical literacy in the third space. *Reading research quarterly*, *43*(2), 148-164. https://doi.org/10.1598/RRQ.43.2.3

[29] (text in Greek) Mellos, A. (2025). Utilisation of the Viber application in distance collaborative projects: A case study of students at the Hellenic Open University [Master's thesis, Hellenic Open University]. https://apothesis.eap.gr/archive/item/216978?lang=en [Access Date: 01/01/2026]

[30] Annamalai, N. (2019). Using WhatsApp to extend learning in a blended classroom environment. *Teaching English with Technology*, *19*(1), 3-20. https://www.ceeol.com/search/article-detail?id=737362 [Access Date: 01/01/2026]

[31] Dahdal, S. (2020). Using the WhatsApp social media application for active learning. *Journal of Educational Technology Systems*, *49*(2), 239-249. https://doi.org/10.1177/0047239520928307

[32] Tang, Y., & Hew, K. F. (2022). Effects of using mobile instant messaging on student behavioral, emotional, and cognitive engagement: a quasi-experimental study. *International Journal of Educational Technology in Higher Education*, *19*(1), 3. https://doi.org/10.1186/s41239-021-00306-6

[33] Palaroan, R. M., Gimeno, A. R., Biñas, G. C., Boloron, R. A., Budomo, X. M., De Guia, L. C., ... & Heidarzadegan, N. (2023). Connecting Beyond the Classroom: Use of Viber as a Support Tool for Enhancing Essay Writing Skills and Online Language Learning Engagement among Students. *World Journal of English Language*, *13*(3), 265. https://econpapers.repec.org/article/jfrwjel11/v_3a13_3ay_3a2023_3ai_3a3_3ap_3a265.htm [Access Date: 01/01/2026]

[34] Solano, G. L., & Miller, J. (2025). Improving Communication and Collaboration: How Using a Communication App Improves the Online Student Learning Experience. *Journal of Educational Research and Practice*, *15*(1), 20. https://scholarworks.waldenu.edu/jerap/vol15/iss1/20/

[35] Lowenthal, P. R., & Dunlap, J. C. (2018). *Investigating students' perceptions of instructional uses of social media*. Distance Education, 39(2), 205–220. https://doi.org/10.1080/01587919.2018.1476844

[36] Belt, E. S., & Lowenthal, P. R. (2021). Video use in online and blended courses: A qualitative synthesis. *Distance Education*, *42*(3), 410-440. https://doi.org/10.1080/01587919.2021.1954882

**Contribution of Individual Authors to the Creation of a Scientific Article (Ghostwriting Policy)**

The authors equally contributed in the present research, at all stages from the formulation of the problem to the final findings and solution.

**Sources of Funding for Research Presented in a Scientific Article or Scientific Article Itself**

No funding was received for conducting this study.

**Conflicts of Interest**

The authors have no conflicts of interest to declare that are relevant to the content of this article.